\documentclass{knawproc}

\begin{document}

\begin{opening}

\title{Old stellar populations in distant radio galaxies}

\author{James S. Dunlop}
\addresses{Institute for Astronomy, University of Edinburgh, Royal
Observatory, Blackford Hill, Edinburgh EH9 3HJ, UK}
\runningtitle{Old stellar populations}
\runningauthor{J.S.Dunlop}

\end{opening}

%%%%%%%%%%%%%%%%%%%%%%%%%%%%%%%%%%%%%%%%%%%%%%%%%%%%%%%%%%%%%%%%%%%%%%%%%%%%%%%

\begin{abstract}
I describe the current status of our attempts to determine the age of the
oldest known stellar populations at high redshift, in the red mJy radio
galaxies 53W091 ($z = 1.55$) and 53W069 ($z = 1.43$). During the past
year the original conclusion of Dunlop et al. (1996) - that 53W091 is $>3$
Gyr old - has been questioned from two, basically orthogonal directions. 
First, reports that the near-infrared light from 53W091 is highly
polarized have cast some doubt on whether its red colour is genuinely due
to an old population of stars. Second, assuming that all the light is indeed
due to stars, it has been claimed that 53W091 is in fact only 1-2 Gyr
old. Here I present a preliminary analysis of new infrared polarimetric
observations of 53W091 which show that the first of these criticisms can
be rejected with very high confidence. I then explore why different
modellers have derived different ages for 53W091, and present new model
fits to the spectrum of 53W069 which demonstrate that different spectral
synthesis codes are certainly in good agreement that this galaxy 
is 3-4 Gyr old. Finally I present a preliminary analysis of the 
morphologies and scale-lengths of 53W091 and 53W069 as derived from  new 
I-band WFPC2 HST images, and compare the results with those for 3CR 
galaxies at comparable redshifts. I conclude that the scalelengths and 
luminosities of radio galaxies at $z \simeq 1.5$ appear to scale together 
as would be predicted from the Kormendy 
relation for low-redshift elliptical galaxies.
\end{abstract}

%%%%%%%%%%%%%%%%%%%%%%%%%%%%%%%%%%%%%%%%%%%%%%%%%%%%%%%%%%%%%%%%%%%%%%%%%%%%%%%

\section{Introduction}

\subsection{Cosmic star formation and radio galaxies}

The study of `normal' star-forming galaxies at $z > 2$ has developed into
a booming astronomical industry over the past 18 months ({\it e.g.}
Steidel et al. 1996; Giavalisco et al. 1996; Steidel et al. 1997).
When coupled with the results from complete redshift surveys reaching $z
\simeq 1$ ({\it e.g.} Lilly et al. 1995) such studies have been used to
produce the first estimates of the star formation history of the universe
(Madau 1997) as shown in Figure 1. However, a number of lines of evidence
suggest that this picture is, perhaps unsurprisingly, biased and incomplete
due, at least in part, to the effect of dust. First, as discussed by
Dunlop (1997), and shown in Figure 1, the evolving radio luminosity density
produced by the radio-loud active-galaxy population traces the ultraviolet 
luminosity density produced by stars rather well out to $z \simeq 1$,
suggesting that, averaged on a large enough scale, both processes 
may simply scale with a global cosmological fueling rate. There is no
real reason to suppose that this should not also apply at higher redshift
and so, since the evolving radio luminosity function is now rather well
determined out to $z \simeq 4$ and is immune to dust obscuration, Dunlop
(1997) argued that the radio based curves shown in Figure 1 should be
regarded as our current `best bet' as to the true star-formation history of
the Universe. Second, recent attempts to determine the correction due to
dust obscuration which should be applied to the high redshift data points
in Figure 1 do indeed seem to be sufficient to make them consistent with
the radio-based curves; Pettini et al. (1997) suggest they be raised by a
factor of $\simeq 3$ while Heckman (this meeting) suggests a factor of 
$\simeq 7$ may be appropriate, broadly consistent with the results of 
Sawicki et al. (1997).
Finally, it remains unclear whether Figure 1 contains any information at
all about the formation of massive elliptical galaxies. Certainly out to
$z \simeq 1$ essentially all of the ten-fold increase in star-formation
activity occurs in the irregular/spiral population. Furthermore 
a number of lines of evidence continue to indicate that
the majority of stars in massive ellipticals formed in a relatively 
short-lived (and potentially dust-enshrouded) burst at high redshift 
({\it e.g.} Bower et al. 1992, Zepf \& Silk 1996). The key question is how 
high a redshift?

\begin{figure} 
  \vspace{21.2pc} 
\includegraphics{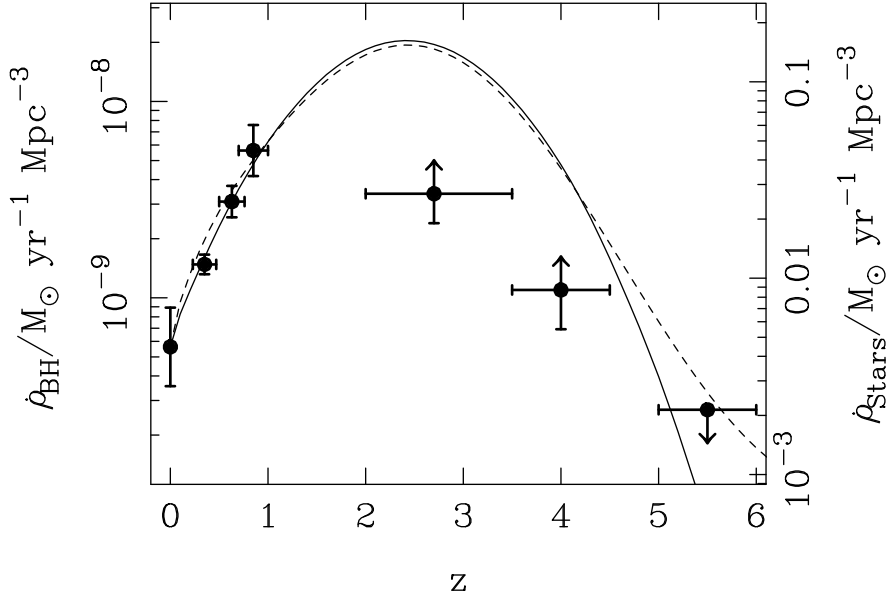} 
\caption[]{A comparison of the redshift dependence of 
the rate at which mass is consumed by black 
holes at the centre of giant elliptical galaxies (and turned into 
radio luminosity from AGN; curves 
and left-hand axis) with the rate at which mass is converted into stars 
per Mpc$^{3}$ (and turned into
UV-light from primarily disc/irregular galaxies; data points and 
right-hand axis). 
The solid and dashed curves
are the luminosity-weighted integrals of the pure luminosity evolution
(PLE)
and luminosity/density evolution (LDE) evolving radio
luminosity functions described by Dunlop \& Peacock (1990), 
converted into black hole 
mass consumption rate per Mpc$^3$ assuming an efficiency of $\simeq 1$\%. 
The data-points
indicating the star-formation history of the Universe are taken from
Madau (1997), and are themselves derived from a low-redshift H$\alpha$
survey ($z \simeq 0$), the Canada France Redshift Survey ($z < 1$), and the 
number of colour-selected `U-dropout',
`B-dropout' and (lack of) `V-dropout' galaxies in the Hubble Deep Field.
The upward pointing arrows indicate the fact that the assessment of
star-formation rates based on Lyman-limit galaxies is liable to be
under-estimated due to the effects of dust.}
\end{figure} 

Studies of radio galaxies remain one of the most effective ways of
addressing this question. One of the cleanest results in extra-galactic
astronomy is that all powerful ($P > 10^{24} {\rm W Hz^{-1} sr^{-1}}$)
radio sources in the present-day universe are hosted by giant
ellipticals. It is thus not unreasonable to assume that high-redshift
radio sources also reside in ellipticals or their progenitors. Thus while 
Figure 1 indicates that the {\it radio activity} produced by these objects
traces overall star-formation history, their host galaxies belong to the
special subset of giant ellipticals. Furthermore, radio selection is 
less biased towards actively star-forming objects than is optical
selection. The study of the reddest radio galaxies at high-redshift
should  thus provide an efficient and powerful means of determining the
formation epoch of massive ellipticals, provided one can establish that
their optical-infrared properties are not significantly distorted by the
presence of an active nucleus.

\subsection{Lessons learned from 3C65}

The first radio galaxies studied in detail at optical-infrared
wavelengths were selected from the 3CR sample ({\it e.g.} 
Lilly \& Longair 1984),
and the reddest 3CR galaxy at $z \simeq 1$ is 3C65. This galaxy has been
been the subject of several rather detailed studies, the results of 
which serve to highlight two points of particular relevance to attempts
to determine the age of high-redshift radio galaxies in general.
First, while Stockton et al. (1995) showed that this
object possesses a strong 4000\AA\ break consistent with a well-evolved
population, the Keck spectrum of this most passive of 3CR galaxies at $z
\simeq 1$ still contains many emission lines and remains dominated
by AGN-related emission shortward of $\lambda_{rest} \simeq 3000$\AA.
Second, attempts to date 3C65 have 
revealed that the spectrophotometric models
of Bruzual \& Charlot (Chambers \& Charlot 1990; Bruzual \& Charlot 
1993;1997) yield very different ages
when applied to individual spectral features as compared to
optical-infared colours; fitting the 4000\AA\ break with the models of
Bruzual \& Charlot led Stockton et al. (1995) to conclude in favour
of an age of 4 Gyr for 3C65, whereas Chambers \& Charlot (1990) 
had previously derived an age of only 1.7 Gyr from its $R-K$ colour.

\subsection{Red mJy radio galaxies}

By selecting radio galaxies at mJy flux levels we have shown that it is
possible to find examples of well-evolved galaxies at $z \simeq 1.5$
whose near-ultraviolet spectrum appears essentially uncontaminated by the
direct or indirect effect of AGN activity. The two best examples
studied to date are 53W091 at $z = 1.552$ 
(Dunlop et al. 1996; Spinrad et al. 1997) and 
and 53W069 at $z = 1.432$ (Dey et al. 1997). Keck spectroscopy of these
objects has yielded the first detection of stellar absorption features
from `old' stars at $z \ge 1.5$ and thus the first `reliable' age-dating
of evolved high-redshift objects. However, during the past year the 
conclusions of Dunlop et al. (1996) and Spinrad et al. (1997) regarding
the age of 53W091 have been challenged from two rather different
directions. In the next two sections I therefore assess
the validity of these criticisms.

\section{Polarized near-infrared light in 53W091 ?}

It has recently been claimed that new 
polarimetric observations of 53W091 show that it is strongly polarized at 
near-infrared wavelengths with $p \simeq 40$\% (Chambers, priv. comm.).
The full details of this
observation have yet to appear in the literature, but nevertheless we
felt that it was important to attempt an independent check of the
validity of this result because of its obviously important implications
for the interpretation of the red colour of 53W091.

Therefore, on the nights of 17/18 August 1997 we observed 53W091 at $K$
with the dual-beam polarimeter IRPOL2 on UKIRT for a total integration
time of 6 hours. Full details of these observations will be presented by
Leyshon et al. (1998) while the data reduction method is described in detail
by Leyshon \& Eales (1997). The result of this observation is that 
the (debiased) percentage polarization of the $K$-band light from 53W091
is $p = 0.31$\%, with a 1-$\sigma$ confidence range of $0 < p < 9.6$\%.
We therefore find no evidence of significant polarization in the infrared
light from 53W091, consistent with our original conclusion that the light
from 53W091 is dominated by an old stellar population from
$\lambda_{rest} \simeq 2000$\AA\ to $\lambda_{rest} \simeq 1 {\rm \mu m}$.

\begin{figure}
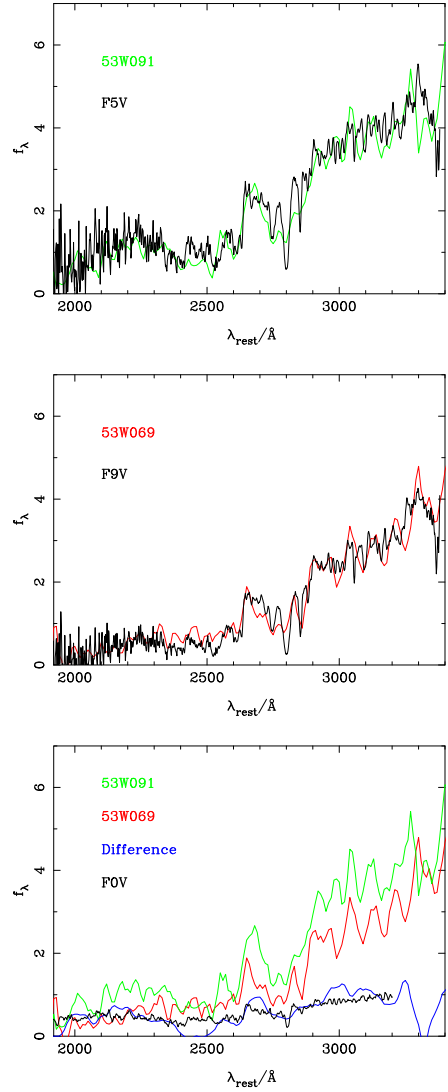
 
\vspace{33pc} 
\includegraphics{knaw_fig2a.eps}
\includegraphics{knaw_fig2b.eps}
\includegraphics{knaw_fig2c.eps}
\caption[]{\small Top Panel: Galaxy rest-frame Keck spectrum of 
53W091 compared with an average F5V IUE spectrum.
Middle Panel: Galaxy rest-frame Keck spectrum of 53W069 compared with an average F9V IUE 
spectrum. Bottom Panel: Comparison of the properly scaled rest-frame UV
spectra of 53W091 and 53W069 with a smoothed version of the difference
spectrum compared with an average F0V IUE spectrum.}
\end{figure}

\section{Age controversy}

\subsection{Comparison of 53W091, 53W069 and stars}

In Figure 2 I show the rest-frame ultraviolet spectra of 53W091 and 53W069
as derived from our Keck spectroscopy, 
overlaid with the average IUE spectrum of the spectral type of star which 
best fits the near-ultraviolet spectral energy distribution of each galaxy.
53W091 is essentially indistinguishable from an F5V star, while 53W069 is
even redder, and is best described by an F9V star. The redder ultraviolet SED of
53W069 is consistent with the fact that it is 0.5 magnitudes fainter in
$R$ than 53W091 while having the same $K$ magnitude (18.7 in a 4 arcsec
aperture). The spectra in Fig. 2 have been scaled to reflect this 
0.5 mag difference in the $R$-band, and are compared 
directly in the bottom panel. This
plot illustrates rather graphically how many of the features in
the spectra of these galaxies are real, rather than due to noise or poor
sky subtraction (the galaxies having of course been shifted from different
redshifts). Also shown is a smoothed version of the difference between
the two SEDs, which is rather flat (in $f_{\lambda}$). The origin of this
additional low-level blue component in 53W091 is unclear. It may or may
not be starlight, but as shown it
is broadly consistent with the SED of an F0V star. Shortward of 2300\AA\
it appears to contribute approximately half of the emission from 53W091,
a fact of importance when considering some of the recent controversy over
the age of this galaxy (see next section).

\subsection{Spectral Features}

\begin{table}[t]
\caption{The strengths of the near-UV spectral features in the radio
galaxies 53W091 and 53W069 as defined by Fanelli et al. (1992), and
quantified in terms of the spectral type of solar-metallicity star which
displays comparable feature strength (as determined from average 
IUE spectra by Fanelli et al.). In terms of typical feature strength,
both galaxies are essentially indistinguishable from the Sun.}

\vspace{0.4cm}
\begin{center}
\begin{tabular}{|c|c|c|}
\hline
{\bf Feature} & {\bf 53W091}  & {\bf 53W069}\\
\hline
2609/2660\AA\ break &  $G0 \rightarrow G5$ & $G0 \rightarrow G5$ \\ 
2828/2921\AA\ break &  $G0 \rightarrow G5$ & $G0 \rightarrow G5$ \\
FeII 2402\AA        &  $F8 \rightarrow G9$ & $G0 \rightarrow G5$ \\
FeII 2609\AA        &  $F8 \rightarrow G9$ & $F8 \rightarrow G9$ \\
MgII 2800\AA        &  $A0 \rightarrow A2$ & $A5 \rightarrow A8$ \\
MgI  2852\AA        &  $A9 \rightarrow F3$ & $G0 \rightarrow G5$ \\
FeI 3000\AA         &  $G0 \rightarrow G5$ & $G6 \rightarrow G9$ \\
BL 3096\AA          &  $G0 \rightarrow G5$ & $G0 \rightarrow G5$ \\
Mg Wide             &  $G0 \rightarrow G5$ & $G6 \rightarrow G9$ \\
\hline
\end{tabular}
\end{center}
\end{table}

The shape of the ultraviolet SED is potentially sensitive to both
low-level AGN contamination and to reddening by dust. However, the
strengths of individual spectral features defined over a relatively short
baseline in wavelength should be more robust. 
It is therefore useful to compare the
strengths of the spectral features (absorption lines and breaks) in
53W091 and 53W069 with the strengths of the same features in stars of
different spectral class as determined from their IUE spectra.

For 53W091 this comparison was performed by Spinrad et al. (1997) for the
two prominent spectral breaks at 2640\AA\ and 2900\AA. Here I have
extended the measurement/comparison to include all of the most prominent
features in the IUE spectra of stars as defined by Fanelli et al. (1992)
and have also performed the same analysis for 53W069.

The results are presented in Table 1, where the strength of each feature
has been quantified in terms of the range of spectral classes which
display features of comparable strength (Fanelli et al. 1992). 

This comparison demonstrates that, in terms of spectral feature strength, both
galaxies are consistent with the spectrum of a G0V-G5V star. Only the MgII
absorption line is too weak to be consistent with such a spectral
classification, but this line is particularly likely to be weakened by
some emission in a radio galaxy. The similarity of the strengths of the
features in 53W091 and 53W069 strengthens the conclusion of the previous
subsection that the ultraviolet SED of 53W091 only differs from that of
53W069 through the presence of an additional low-level blue component.
These results also indicate that 53W069 is a particularly `clean' object,
in which the spectral type of star which best describes the strength of
its spectral features is consistent with the spectral type which
best fits the overall shape of its ultraviolet SED.

This conclusion is obviously independent of any spectral synthesis
modelling, and helps to clarify the basic question which spectral
modelling must aim to answer - {\it i.e.} how quickly can the spectrum of
a galaxy evolve to be indistingishable from that of a G0 star (or, within
the errors, identical to that of the Sun)?

\subsection{Model Fitting}

Recently Bruzual \& Magris (1997) have reported that the spectral energy
distribution of 53W091 is bluer than that of M32 and, using the models of
Bruzual \& Charlot, that it can be reproduced by an 
instantaneous starburst with on age of only 1-2 Gyr. The former
conclusion is basically in agreement with the comparison of 53W091 and
M32 performed by Dunlop et al. (1996), but the latter conclusion is not.

In order to clarify the origin of this discrepancy, I have performed
properly weighted chi-squared fits to the ultra-violet SEDs of 53W091,
M32 and 53W069 using the models of Bruzual \& Charlot and the new revised
models of Jimenez et al. (1997) (based on new isochrone calculations
performed on the Cray T3D at Edinburgh). The results are listed in Table 2,
where for 53W091 and 53W069 I also give the ages derived from $R-K$
colour and the strength of the breaks at 2640\AA\ and 2900\AA.

\begin{table}[t]
\caption{A comparison of age estimates for the stellar populations in
53W091, 53W069 and M32 as derived from the instantaneous burst models of
Bruzual \& Charlot (B\&C) and  Jimenez et al. (1997) (J97), 
when used to fit different spectral indicators of age.}

\vspace{0.4cm}
\begin{center}
\begin{tabular}{|l|c|c|c|c|c|c|}
\hline
  & \multicolumn{2}{c|}{\bf 53W091} & \multicolumn{2}{c|}{\bf 53W069} &
\multicolumn{2}{c|}{\bf M32} \\
\hline
{\bf Feature} & {\bf B\&C}  & {\bf J97} & {\bf B\&C}  & 
{\bf J97} & {\bf B\&C}  & {\bf J97}\\
\hline
UV-SED   & 2.5 Gyr & 3.5 Gyr & 3.3 Gyr & 4.3 Gyr & 2.8 Gyr & 3.5 Gyr\\
$R-K$    & 1.4 Gyr & 3.0 Gyr & 1.6 Gyr & 4.0 Gyr & & \\
2640\AA\ & 5.0 Gyr & 4.0 Gyr & 5.0 Gyr & 4.0 Gyr & & \\
2900\AA\ & 4.5 Gyr & 4.0 Gyr & 4.5 Gyr & 4.5 Gyr & & \\
\hline
\end{tabular}
\end{center}
\end{table}

This table helps to clarify a number of important points. First, I should
emphasize that the Bruzual \& Charlot models to which I had access
when performing these fits was not the latest (1997) version used by
Bruzual \& Magris (1997). However, the age I have derived by fitting the
Bruzual \& Charlot models to the
ultraviolet SED of M32 is 2.8 Gyr, in excellent agreement with that
reported by Bruzual \& Magris (1997), indicating that in the age-range
and spectral-range of interest the models have changed very little.

Second, columns 2 and 4 re-emphasize that, as already discussed in
section 1 in the context of 3C65, the Bruzual \& Charlot models seem to
be internally inconsistent in the sense that they are capable of
reproducing very red $R-K$ colours at a much younger age than they can
reproduce the ultraviolet SED or the spectral breaks in 53W091 and 53W069.
However, it is well-known that the correct prediction of $R-K$ colours at
young ages depends crucially on the treatment of the late stages of
evolution, whereas prediction of the UV-spectrum at such times depends 
mainly on the correct prediction of the main sequence turn-off point.
As discussed by Jimenez et al. (1997), it appears that the more
consistent $R-K$ ages produced by the Jimenez et al. (1997) models (see columns
3 and 5) result from a more sophisticated treatment of mass-loss on the
red giant branch which prevents the production of a very red horizontal
branch, or such a red and luminous asymptotic branch. Whatever the
precise origin of the apparently 
excessively red colours produced at young ages 
by the Bruzual \& Charlot models, we believe it is much safer to focus on
ages determined from the fits to the UV-SED and spectral features, which
are less sensitive to our precise knowledge of the latter stages of
stellar evolution. Indeed, not to do so it tantamount to throwing away
the new, more robust information which can be gleaned from the
spectroscopy (note that Bruzual \& Magris 1997 cite the ability of
their models to reproduce the entire SED of M32 at a single age (2.8 Gyr) as
a success. However, this could in fact be regarded as problematic since
the spectrum of a present-day elliptical is expected to be consist of a
number of components of different ages which will dominate at different
wavelengths. Arguably a more important test of the internal 
consistency of such
models is their ability to reproduce the entire SED of a high-redshift
galaxy whose population is more likely to be coeval).
 
\begin{figure} 
\vspace{35pc} 
\includegraphics{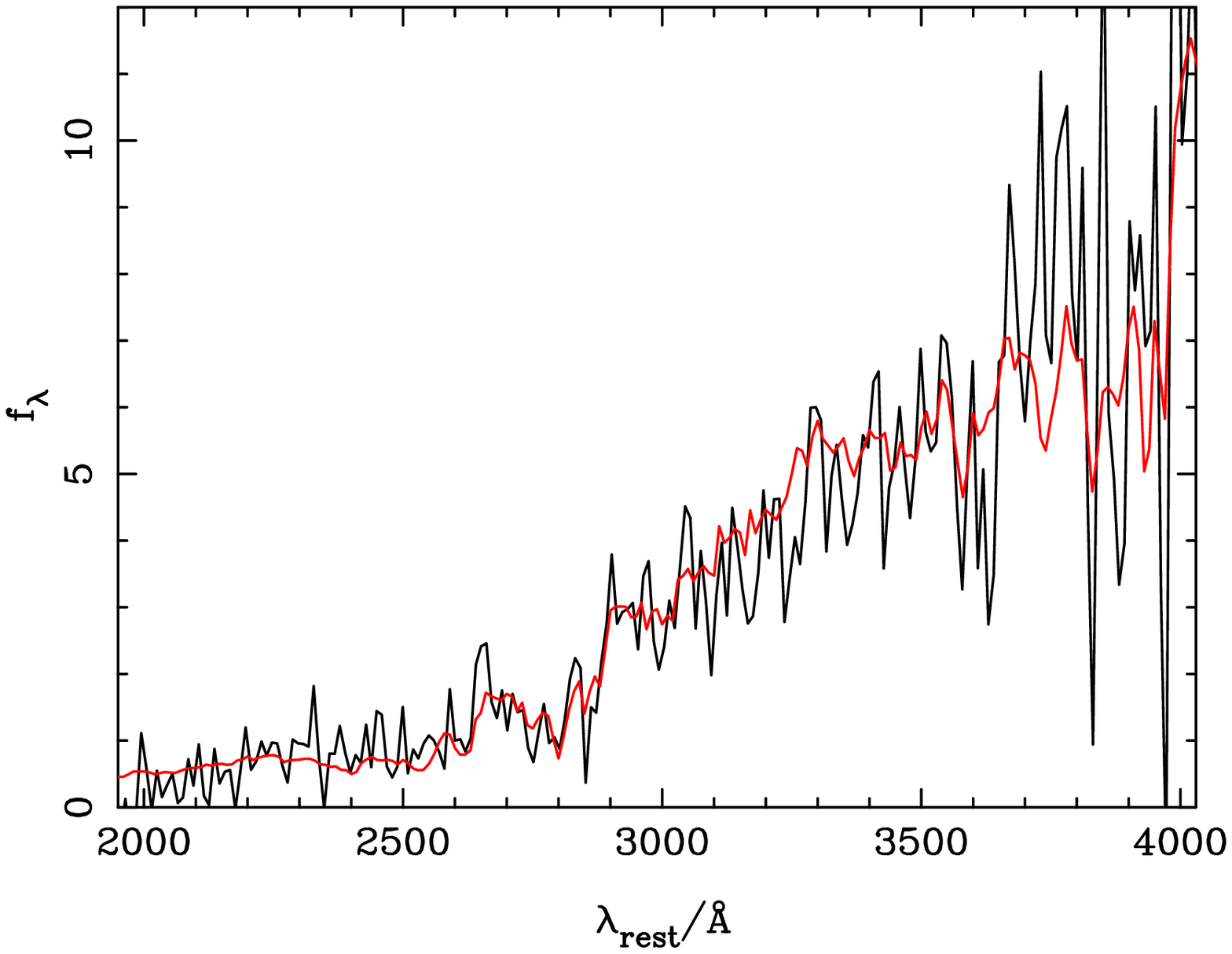}
\includegraphics{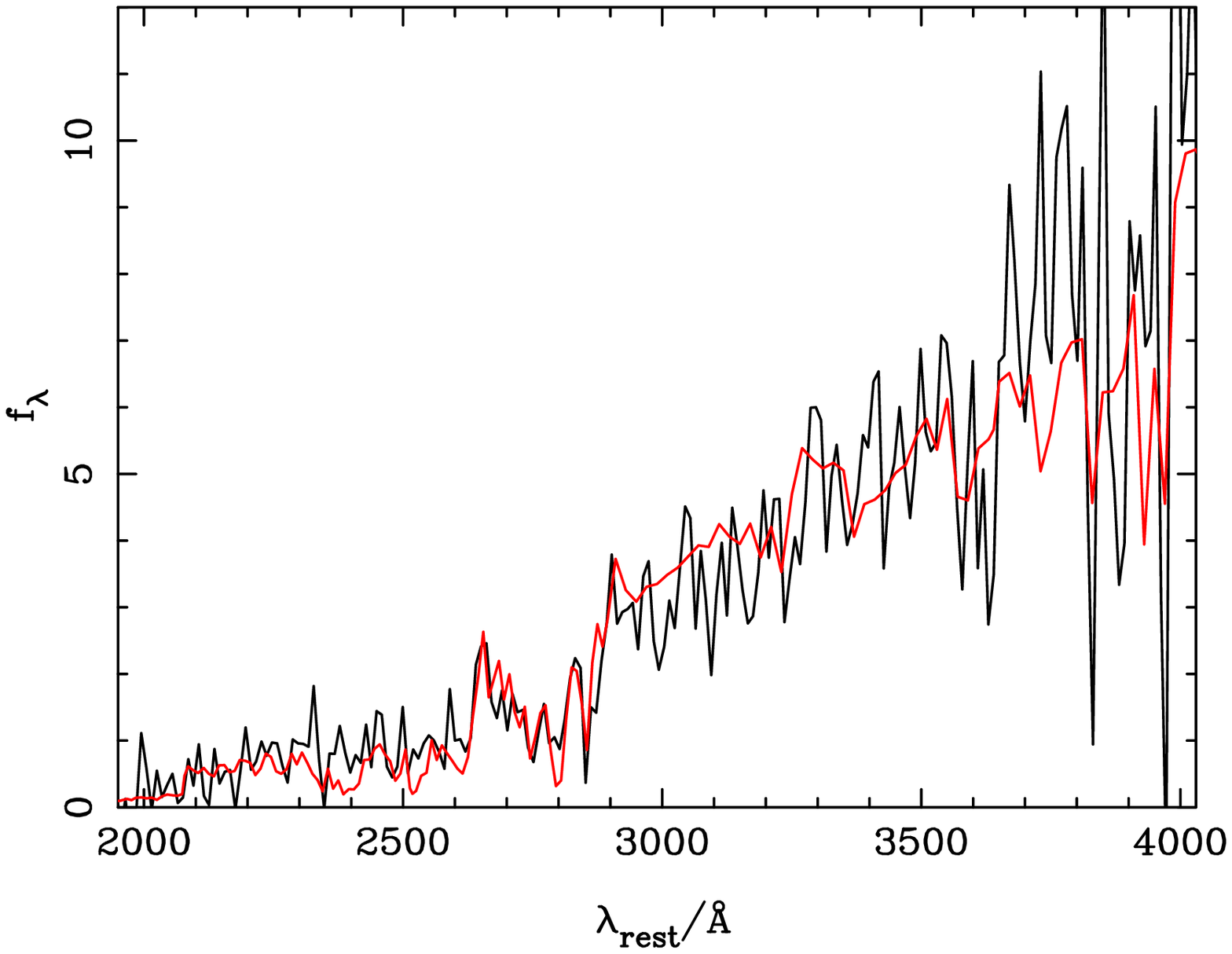}
\caption[]{\small Upper panel: The spectrum of 53W069 overlaid with the 
best fitting Bruzual-Charlot model, which has an age of 3.25 Gyr.
Lower panel: The spectrum of 53W069 overlaid with the best fitting model
from Jimenez et al. (1997), which has an age of 4.25 Gyr.}
\end{figure}

Focussing, therefore, on UV dating it is clear that both sets of models
agree that 53W091 is marginally bluer/younger than M32 but that 53W069 
is redder/older.
Also, the ages derived from the Bruzual \& Charlot models are, in each
case, $\simeq 0.9$ Gyr younger than those derived from the new Jimenez
models. Tracing the origin of this second 
discrepancy is harder and needs detailed investigation, but we suggest
that it may be primarily due to the difference between the isochrones
produced by the Padova tracks (used by Bruzual \& Charlot) 
and those of Jimenez et al. (1997). The former are typically 100K cooler
than the latter at the epochs of interest, which translates to a
difference in inferred age of up to 1 Gyr. Jimenez et al. (1997)
present evidence based on Hipparcos data that the Padova tracks are in
fact too red.

Whatever the precise origin of this disagreement, it is in any case not
so severe as to affect our basic conclusion.
If one focusses on the results for 53W069 then, ignoring the anomalously
young $R-K$ age produced by the Bruzual \& Charlot models, both models
are basically in good agreement that the overall shape of the UV SED and
the strengths of the main spectral features are consistent with an age in
the range 3.5 $\rightarrow$ 4.5 Gyr. In Figure 3 I compare the spectrum
of 53W069 with both the best-fitting Bruzual \& Charlot model (3.3 Gyr)
and the best-fitting Jimenez et al. (1997) model (4.3 Gyr).

Finally, I note that while Table 2 indicates that while on the basis of
fits to the its ultraviolet SED 53W091 appears to be 
about a Gyr younger than 53W069, dating based on spectral breaks yields
an age more similar to that of 53W069, again consistent with the
conclusion that 53W069 is a cleaner example of a genuinely passively
evolving galaxy.

In summary, the new data on 53W069 coupled with this more detailed
analysis appears to basically re-affirm and 
strengthen the original conclusion of Dunlop
et al. (1996) that, assuming solar metallicity, $3-4$ Gyr have elapsed
since the cessation of the major burst of star-formation in these
galaxies at $z \simeq 1.5$.

\section{HST observations}

\subsection{Morphology and scalelengths}

Spectroscopically, 53W091 and 53W069 thus appear to be the best known
examples of old, passively-evolving elliptical galaxies at redshifts as
high as $z \simeq 1.5$. It is therefore obviously of 
interest to determine whether this is also true morphologically, and to
attempt to determine their scalelengths. To address this issue we were
awarded 9 orbits of Cycle 7 HST time to obtain WFPC2 and NICMOS 
images of both galaxies below and above the 4000\AA\ break 
(through the F814W and F110W filter respectively). Full details of these
observations and our results will be presented in Peacock et al. (1998b),
but here I briefly discuss the results of a preliminary analysis of the
WFPC2 data. 

I have used a modified version of the 2-dimensional fitting code
developed by Taylor et al. (1996) to determine whether both galaxies are
better described by a de Vaucouleurs $r^{1/4}$ law, or by an exponential
disc. As illustrated by the 1-dimensional luminosity profiles of 53W069
shown in Figure 4, both galaxies are perfectly consistent with a de
Vaucouleurs law and inconsistent with a disc. Moreover, the fitted
observed 
half-light radii of both galaxies are very similar; $r_e = 0.43$ arcsec
for 53W069, and $r_e = 0.42$ arcsec for 53W091. Assuming $\Omega_0 = 1$
and $H_0 = 50 {\rm km s^{-1} Mpc^{-1}}$ this implies a physical 
half-light radius of $r_e \simeq 4$ kpc for both objects.

\begin{figure}
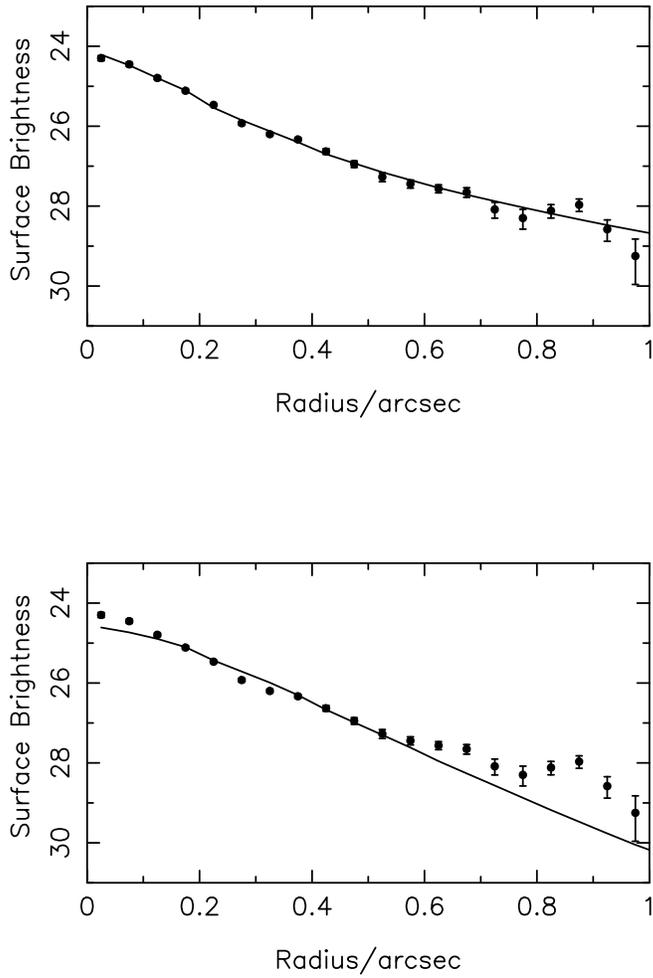
 
\vspace{35pc} 
\includegraphics{knaw_fig4a.eps}
\includegraphics{knaw_fig4b.eps}
\caption[]{\small Upper panel: The high-quality fit to the 
$I$-band (F814W) luminosity profile of 53W069 which is produced using a 
de Vaucouleurs law convolved with the HST F814W point spread
function. Lower panel: The inadequate fit to the same data 
which is provided by the best-fit 
disc model convolved with the HST F814W point spread
function.}
\end{figure}

\subsection{K-z diagram and the Kormendy relation at high z}

These first reliable scale-length detections for weak radio galaxies at
high redshift, when coupled with the HST-based 
scalelength determinations for the 3CR galaxies of Best et al. (1997)
allow a first attempt to determine whether high-redshift radio galaxies
still follow the Kormendy relation for giant ellipticals, as they do at
low redshift (Taylor et al. 1996).
 
Both 53W091 and 53W069 have basically identical $K$-band magnitudes of
18.3 within an 8 arcsec aperture. Reference to the most recent version of
the $K-z$ diagram for radio galaxies as presented by Eales et al. 
(1997) shows that these 2 galaxies lie $\simeq 0.7-0.8$ mag faintward
of the mean $K-z$ relation for the 3CR galaxies at comparable redshift.
The more radio-luminous 3CR galaxies are therefore on average brighter,
but if this were generally due to an enhanced nuclear contribution as
suggested by Eales et al.  (1997), they would be expected to be 
more nucleated. In fact,
the opposite is true, they are larger (Best et al. 1997) 
providing further circumstantial evidence that their infrared
emission is indeed dominated by starlight.

This baseline in magnitude allows us to test, for the first time, 
whether the magnitudes and scalelengths of radio galaxies at $z > 1$ 
are consistent with the Kormendy relation. The answer is that they are.
Best et al. (1997) have determined that the average
half-light radii of the 3CR radio galaxies at $z > 1$ is $r_e \simeq 10$
kpc (note that Best et al. (1997) were not able to constrain the
scalelengths of the 3CR galaxies to sufficient accuracy to prove that
they follow the Kormendy relation, but did at least manage to 
show that their
average position on the $\mu_e-r_e$ diagram was consistent with a
passively evolved version of the low-redshift Kormendy relation for
ellipticals). If we simply ask what scalelength would then be predicted
for an elliptical 0.8 mag fainter, the answer is 4 kpc, exactly
consistent with that derived from our new HST images of 53W091 and 53W069.

\section{Conclusion}

Our Keck and HST data on the mJy radio galaxies 53W091 and 53W069
basically re-affirm the usefulness of high-redshift radio galaxies as
probes of the evolution of massive ellipticals, although it is clear that
one needs to avoid the most radio-luminous sources to obtain an
uncontaminated view of the underlying stellar population in the
rest-frame optical-ultraviolet. Our new infrared polarimetric observations 
of 53W091 coupled with detailed modelling of the Keck spectra of 53W091 and
especially 53W069 allow the original conclusions of Dunlop et al. (1996)
to now be re-affirmed with increased confidence - the observed
optical-infrared light of these galaxies is dominated by an evolved
stellar population with a minimum age of $3-4$ Gyr. Analysis of the
$I$-band WFPC2 HST images of both galaxies confirm that, morphologically, 
they are relaxed elliptical galaxies with a scalelength $r_e \simeq 4$
kpc, exactly as would be predicted by the Kormendy relation for
elliptical galaxies given that they are 0.8 mag fainter at $K$ than the
3CR galaxies which have an average scalelength of $r_e \simeq 10$kpc.
Our results therefore still point towards a very high formation
redshift for massive elliptical galaxies, and continue to place a severe strain
on an Einstein de-Sitter cosmology. However, as discussed elsewhere in
this volume by John Peacock (see also Peacock et al. 1998a), provided one
adopts a present-day age of the universe of $\simeq 14$ Gyr, it is perfectly
reasonable to expect a small number of $L^*$ ellipticals to be as old as 
3-4 Gyr by $z \simeq 1.5$. The next important step will be to determine the
true space density of comparably old objects from infrared-based surveys.

\begin{acknow}

I gratefully acknowledge the contributions of my collaborators 
John Peacock, Raul Jimenez, Hy Spinrad, Arjun Dey, Daniel Stern, Rogier
Windhorst, Steve Eales, Gareth Leyshon, Ian Waddington and Ross McLure.
\end{acknow}

\end{document}